# AN APPROACH FOR A BUSINESS-DRIVEN CLOUD-COMPLIANCE ANALYSIS COVERING PUBLIC SECTOR PROCESS IMPROVEMENT REQUIREMENTS

Sachar Paulus[1] and Ute Riemann[2]

[1]Department of Economics, Brandenburg University of Applied Sciences, Brandenburg, Germany

[2]SAP Deutschland AG & Co KG, Walldorf, Germany

**ABSTRACT**

*The need for process improvement is an important target that does affect as well the government processes. Specifically in the public sector there are specific challenges to face .New technology approaches within government processes such as cloud services are necessary to address these challenges. Following the current discussion of „cloudification"of business processes all processes are considered similar in regards to their usability within the cloud. The truth is, that neither all processes have the same usability for cloud services not do they have the same importance for a specific company.*

*The most comprehensive process within a company is the corporate value chain. In this article one key proposition is to use the corporate value chain as the fundamental structuring backbone for all business process analysis and improvement activities. It is a pre-requisite to identify the core elements of the value chain that are essential for the individual company's business and the root cause for any company success. In this paper we propose to use the company-specific value-creation for the "cloud-affinity" and the "cloud-usability" of a business process in public sector considering the specific challenges of addressing processes in cloud services. Therefor it is necessary to formalize the way the processes with its interdependencies are documented in context of their company-specific value chain (as part of the various deployment- and governance alternatives (e.g. security, compliance, quality, adaptability)). Moreover, it is essential in the public sector to describe in detail the environmental / external restrictions of processes.. With the use of this proposed methodology it becomes relatively easy to identify cloud-suitable processes within the public sector and thus optimize the public companies value generation tightly focused with the use of this new technology.*

## 1. INTRODUCTION

Doing more with less. Delivering better service. Improving collaboration with other agencies. In a nutshell, this is what public services drive today. There is also a constant pressure to do more with less; driving productivity improvements in the public sector is a key success factor for modern e-government. While there are severe challenges, technological developments, including the use of cloud services provide opportunities for innovation that were not previously available.

Innovation in a public sector context can be defined as the creation and implementation of new processes, products, services and methods of delivery which result in significant improvements in the efficiency, effectiveness or quality of outcomes. Innovation is driven by a number of factors, of short, medium and long term range. In the public service the driving imperative for innovation is the need to respond effectively to new and changing government and community expectations in an increasingly complex environment. The benefits of innovation are diverse. It is widely recognized that innovation is crucial to enhanced economic performance, social welfare and environmental sustainability. Innovations can also „improve organizational efficiency; provide higher quality and more timely services to citizens; reduce business transaction costs; and provide new methods of operation. Innovation can enable better performance and drive new directions"[6] .

An innovative activity in the public sector in this sense can be considered in various ways. The three most common ways are:

- Guiding-driven policies
  The role the public sector is to provide target-driven advices to assist the decision making in relation to public services
- Service-oriented policies to support the public services community efficientöy and effectively
- Innovative processes to improve productivity in public services with the overall target of cost reduction

Having these challenges and major changes in mind the need of new processes, structures and system support in public services become obvious. The modernization in regards to a people-centric organization is right at the beginning to emerge. In addition the cost factor drives the enhancement of operational efficiency even further. Recently, cloud computing has captured significant attention in the public sector due to new system functionalities and a stustainable and trustworthy operating model.

Certainly the use of cloud services are an option to consider – however the use of cloud services shall be analysed in a comprehensive way making sure that any compliance issues are identified in advance to proactively consider them. Such a comprehensive analysis regarding the security requirements using cloud services need to cover various dimensions – this is not new. However, most of the currently available analysis frameworks consider system-related, organizational and competency dimensions, and consequently the security requirements are addressed with the „classical", available instruments (e.g. ISO 2700x) accompanied by the burning of a high investment. This paper proposes to go a step further and to address context-independent cloud-based business processes based on the assumption that all involved parties (especially the cloud service provider) need to accept the fact to consider the maximal risk level and consequently to investigate all potentially possible measures.

"The time frames in which the public sector is required to respond are tending to shorten to meet government imperatives, and citizen and stakeholder demands" [6]. At the same time government is looking for, and citizens are demanding, a more holistic or citizen-focused approach to service delivery.

There are commonalities, differences and synergies between private and public sector innovation. „Some aspects of public sector innovation are comparable with, indeed might be almost identical to, aspects of private sector innovation (examples include business process improvements and many aspects of information and communication technologies)" [6]. However, there are other aspects of public sector innovation, particularly those associated with policy innovation, for which governments must bear responsibilities that greatly outweigh those borne by the private sector (examples are national security, counterterrorism and pandemic preparedness).

There is a tremendous pressure on public services to further improve the efficiency and effectiveness of their processes. With the new orientation towards a user-centric service model it is part of their overall strategy and key target[6]. Besides the industry and the specific value chain the cost of compliance leads to the conclusion that the analysis of cloud compliance need to start at the stage of process definition. There are various papers available to analyse the cloud-related value chain. However, these papers consider adjusting the organization towards the cloud-specifics. In other words: the technology of cloud is the key driver to define the corporate value chain.

This paper considers all relevant security criteria alongside the process lifecycle and formulates the process-security management requirements derived from each singe lifecycle phase. We outline the benefit of this approach to formulate, address and meet the relevant security-requirements in an efficient way. In addition, this approach meets the requirements of the Plan-Do-Check-Act-methodology and is fully integrated in a business-motivated evaluation of cloud-affine business processes based considering the corporate value chain. More specifically, we propose to concentrate the effort on the evaluation and choice of cloud-ready business processes (or business process elements), since the „affinity" of business processes for the cloud can be derived from the strategic importance for the value / efficiency creation of the organization and the operational (and compliance) risks that the business process is carrying. A technical basis for this approach is to formalize End-to-End processes of organizations, together with non-functional properties and their impact in different deployment models and governance alternatives. This approach allows to provably choosing those business processes that are specifically suited for cloud usage; the investment into this early compliance activity easily pay back through the avoidance of the implementation of too risky or not enough value-creating business processes in the cloud.

The paper will demonstrate this in the light of public sector business improvement. Cost-cutting, citizen and service quality-orientation, electronic government, and other reform concepts have called for business process improvements of public sector processes . Processes covering eGovernment-related topics have specific challenges to process innovation but as well to accomplish process compliance. The industry of public sector / public sectors organizations in health care, education and social services requires a special understanding on the processes and data that are processed.

We propose

- keeping the corporate value chain as the basis for the appliance of cloud services to get the right selection and added value from the use of cloud. Only if the uniqueness of a corporate value chain is supported by cloud and not changed due to cloud technology the value-add of cloud is given.
- questioning the use of cloud is not an IT question but a joint business and IT question that shall be answered jointly based on a comprehensive assessment result.
- asking and answering the cloud compliance questions not in or after the cloud services have been implemented to run the processes but during the phase of process definition and process identification. The later the relevant questions are answered and appropriate decisions have been made the more impact has the management of any process compliance.

The benefit of the proposed methodology will be demonstrated using examples from the public sector processes. With this approach we believe that the use of cloud services can support each company "uniquely qualification in their markets". We will demonstrate the use and value of our approach with a concrete example of a value chain and on detailed level based on purchasing processes.

## 2. IDENTIFICATION OF A COMPANY-SPECIFIC VALUE CHAIN

In the analysis of the generic value chain in public sector it becomes obvious that thos processes are the key value driving processes that are essential to deliver the required services to the users – in other words: thos processes are key, that have a visible interface to the outside. . The case is that those processes are of special relevance that to deliver a significant portion of the companies value, that are focused on the fulfilment of the customers' needs and consequently create a perceivable customer benefit. The specification of a company's value chain is different for each industry, business model, organization and strategic goals.

Nevertheless the core and supporting processes and the resulting End-to-End processes and overall value chains have shown process standards and process patterns that are almost industry-independent.

A process is a chain of activities tailored to the provisioning of dedicated services. A process is characterized by a services input, service output, a cycle time, handling time and the use of resources ([3] p. 6). The process is an element of a process landscape to gain transparency of the cost driver and the interfaces to other processes.

The starting point to identify the performance indicators are key to a particular company should be the value chain of each company. To assess continuous business process improvement from a process point of view it is necessary to identify the company-specific value-chain. In 1985 Michael Porter described the ways in which a company could organize its activities in order to achieve competitive advantage by making it hard for others to copy [5]. The example of a typical value chain given by Porter included allexternal-facing processes, and in addition their supporting ones. He suggested that „once the value chain had been identified, costs could be assigned to the activities to be able to achieve a cost advantage by reducing the cost of individual value chain activities, or by re-configuring the value chain."[8]. Based on the corporate strategy, the market development as well as the evolving customer preferences we need to give those business processes prominence that are essential for the delivery of goods and services to the customers and/or stakeholders.

The additional step towards the building of a value chain that goes beyond the classical concept of Porter is to segment the value chain in regards to a value-based value chain driven by process value that will be added to the core business of each company and thus strongly linked to the P&L positions. The cost positions and their proportions within selected companies / segments will help to set the right focus and do give an indicator of the importance and value of the area for the business and thus for the significance of the process and its KPIs. Why bother on this value-based segmentation? As said above, it provides the focused context for all work on an efficient and focused value chain management. But most importantly the highest value add is to provide clear definitions, boundaries and dependencies on the most important processes of a company. In a one-step approach the scope is tailored to a manageable but as well most important scope at the same time to let every discussion and activity relate to the core values of a company.

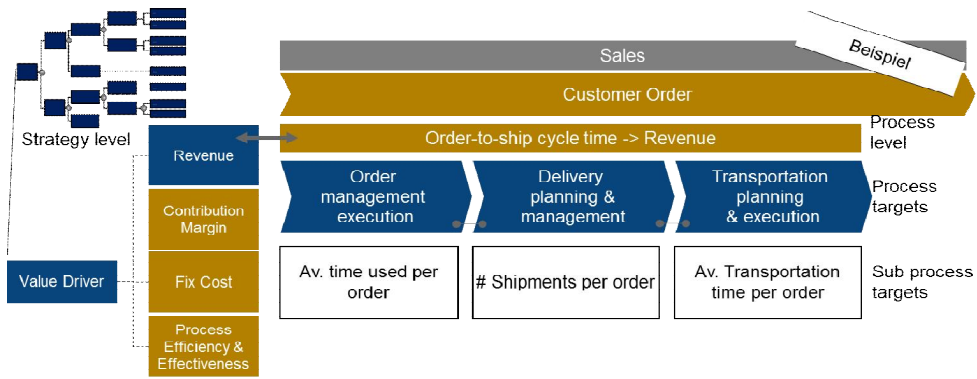

Figure 1: Value drivers to identify the corporate value chain

We will only consider the core processes alongside the value chain on a more detailed level as cloud usability has these processes and the potential change of its deployment is of significant impact and importance. These processes are unique in their design and their value performance. They formulate the unique selling proposition of each company and immediate effect on the operating results and the business excellence.

Especially those processes that do cover more than one business area are important as they are a typical candidate for optimization and thus to simplify integration with the use of cloud services.
To allow a comprehensive company-wide process-analysis we need to consider the supporting processes („Enablers") to a certain extend – however we will not consider them in detail in our paper. Exemplarily we will use the purchasing processes to describe our approach in a practical example.

## 3. CHARACTERISTICS OF THE PUBLIC SECTOR

The most crucial process in public sectors includes administrative processes, customer service and HR and payroll.

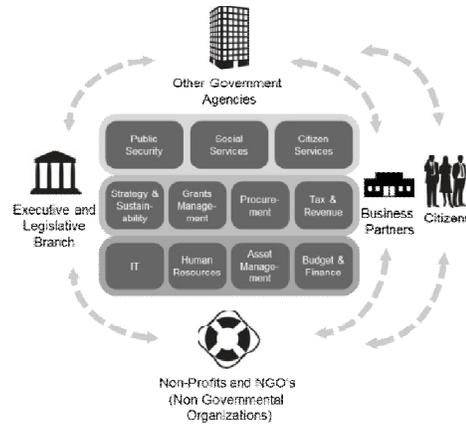

Figure 2: Key Processes in public sector

For the public sector it is the same as for all other industries: Inputs are the triggers for a process; they may be tangible (e.g. a letter from a customer) or intangible (e.g. a need to develop a new service). The output in public sector is mainly taken by the service a person provides it. In other words: the outputs are mainly intangible and closely related to the process performance as well as to the limitations given e.g. by legal constraints. The constraints as well as the perceived level of service needs are the boundaries to be considered when it comes to process optimization as theses constraints need to be considered within the continuous improvement and consistently monitored to be included in the process models.

For this paper we will focus on the generic record to document process that can be applied across the overall public sector. The need to retain records for quality assurance and accountability purposes demands that generic processes be put in place for electronic document and records management. These processes must essentially capture, classify, store, retrieve and use information in collaborative framework (such as case files) followed by their archival and disposal.

The herewith proposed enhancement regarding the usability of cloud services relating to the continuity, the consistency and the governance of business processes while using cloud services. Answering such questions is an important lever for the identification of the inherent costs, potentials and risks and serves as an element to support the decision whether and how to use cloud services for dedicated processes, for the identification of the most appropriate processes and the required establishment of process governance. They should cover topics such as:

• Difficulty in handling increased program and budget demands with scarce resources,
• Department of taxation faces declining revenue streams,
• Agency leadership face concerns over accountability,
• Administrations have too much paperwork and documentation or
• Infrastructure managers have reduced ability to maintain assets and facilities.

The – sometimes expensive- services provided by the public sector are provided with the help of and operational framework consisting of processes and systems in the light of the legal constraints.[7]

While taking advantage of the cloud services we may be able to get support from the generic cloud effects – however we have to consider the costly development of the services and its changes to manage issues and requirements. These issues are not confined to the public sector: the IT systems are necessary to control, administer and enhance the business processes, constrained by regulation – independent from their technical realization. "Thus process compliance is a major topic for business processes in this environment.. Failure to comply is no option at all – this is an even higher restriction as within other businesses"[10].

Therefore we need to have a stable framework to dynamically capture all requirements while allowing the improvement of processes.

## 4. PROCESS IDENTIFICATION TO PROTECT THE VALUE CHAIN

The process identification activity answers the question which business processes of a corporate unit are essential to fulfil the customers and/or other stakeholders' needs based on the business model, the market and the corporate strategy of a company.

The expected optimization with the shift of business processes into the cloud will only bring a positive impact once the profit-relation of the respective („What is the operational effect of the process?") the cost intensity and the fraud indication („What may happen when the process is shifted to the cloud") is evaluated.

End-to end processes for multi-industry value chains such as public sector needs to combining horizontal and industry specific solutions:

- Collaborate better to enhance interagency information and resource sharing
- Adapt better for increased transparency and shifting resource needs
- Decide better by monitoring performance for informed actions for mission success
- Operate better to improve fiscal conditions through a holistic government view

Prior to the use of cloud services and the correlation to the corporate value chain it is important to define meaningful and measurable indicators to allow a careful examination. These indicators need to be as industry-independent as possible to allow on one hand side an application to mostly any industry, market- and competitive situation but on the other hand side need to be as specific as necessary to allow a process appraisal in regards to their cloud service usability.

Following we will consider the following areas of indicators:

1. Result relevance: The importance of a process within the value chain of a company and its value for the operating results in its current status.
2. Cost relevance: It is necessary to check how cost intensive the current process is and which implications will change due to a „cloudification".

3. Security relevance: The dimension covers he assumption of possible frauds due to the use of the process. The indicator can be expressed e.g. in „number of interfaces to 3rd party systems" or „indication in regards to sensitive data". This will give an indication in regards to the potential use of cloud services

Overall, it is necessary to consider End-to-End processes that do cover the whole company (e.g. from customer order to the delivery of goods and/or services). The analysis of sub-processes does not allow identifying fundamental process optimization potentials and process risks. Especially in functional-structured organizations this might lead to misleading assumptions and results, because one single process covers only one single organizational unit with no indication how the overall process works. To get a profound analysis and a sustainable business case it is essential to have an end-to-end focus to analyse the change of the overall process landscape and the corporate risk of a company from a process perspective.

## 5. COMPLIANCE ANALYSIS

We propose that the costs of compliance are related to process but as well to the fact that public sectors have to face different challenges. The challenges in compliance in public sector can be differentiated in three categories (see SAP Thought Leadership, p 5ff):

- Fragmented organization: The disconnection of the organizational units might hinder to implement harmonized, seamlessly linked and/or aligned policies to have the same understanding of risk measurement and the support of regulatory mandates.. Organizational fragmentation is typical within the public sector e confines of its own enterprise and across the extended enterprise.

- Fragemented systems: The information about governing principles and policies, risk measurement, and compliance are still supported in isolated IT systems with no link to each other Local process optimization and implementation of point solutions can further isolate information within systems, resulting in a lack of information integrity and a limited view of enterprise risk.

- Fragmentation due to the local tax regulations: Policies and risks are generally defined and measured at the local level, without proper consideration of their impact on the global, multinational, national, or regional mandates with which an agency must also comply.

In a nutshell: the most significant cloud computing opportunities for the public sector may arise at the multi-agency or all-of-government levels. Around the world, public sector information management is clearly dominated by a "silo" model that sees most government organizations operating largely stand-alone information systems.

If we comprise the thoughts we need to analyse the following topics within the public sector in regards to the use of cloud services:

- How do we run centralized IT services to support multiple departments and agencies?
- How can we better manage IT resources to respond rapidly to business needs?
- How do we control costs through predictable resource allocation?

- How can we provide more consistent and measurable service levels?
- How can we take advantage of outsourcing to reduce expenses?
- How can we reduce server sprawl and increase our efficiency and utilization?

## 6. PROCESS IDENTIFICATION AND LINK TO BENEFIT FROM THE FEATURES OF CLOUD SERVICES

For the identification of the specific value chain we need to first outline the difference of private and public sector and then focus on End-to-End processes. The public sector consists currently of fragmented organizational structures operated by the government. The public sector value chain covers

- Human Capital Management
- Procurement for Public Sector
- Public Sector Accounting
- Social Services and Social Security
- Grants Management
- Tax and Revenue Management
- Organization Management & Support

It needs to be considered that it is not sufficient to analyse the process within the corporate boundaries but to identify the fraud indicators within the company and outside the company to understand the issues that needs to be captured due to the link of processes in the companies outside environment.

### 6.1. Identification of fraud indicators

The focus is on the process related fraud indicators within the public sector. The identification of the process-related fraud indicators shall be therefore driven from the process-typology of the tax and revenue process. As an example the process optimization levers shall be taken to identify the fraud indicators that need to be leveraged. After having identified the fraud indicators the probability of the process step and the fraud level need to be estimated to rate the risk that is available for the corporation. These values (together with the fraud indicators) outline the risk of applying cloud services for a dedicated process within a public service institution.

## 7. EXAMPLE TO APPLY THE CONCEPT AND APPROACH

Following we will outline how the approach and framework is applied to a dedicated process and what the benefits are. Due to complexity reasons we pick up one typical process within the public sector: procurement.

To get the most out of the purchasing functions, the public sector need to gain a consolidated view of the purchasing spent, get efficient purchasing processes and set up a streamlined purchasing organization.

Purchasing is an important lever for the public sector innovation because the spending is quite large and the impact – especially in regards to process efficiency – is quite substantial. The benefits can be summarized as:

- better use of available budget and potential to reduce tax
- increase of transparency – especially of what has been spend
- compliance issues with regulations (e.g. higher degree of assurance that all purchasing across the organization is aligned with the tender regulations)

Even though the realization of innovations is simpler in the purchasing area compared to other areas within the public sector there are specific compliance issues to face. The security concerns can be summarized is:

- Compliance requirements to rely on the agreed procedures of purchasing
- Control of budget (across multiple layers of authority)
- Political objectives need to be considered when streamlining any process, procedure and procurement laws. These regulation frameworks are binding even though they are a constrain for the ability of the public sector to move into a more innovative way of purchasing

Therefore it is essential within the public sector to establish binding rules for all stakeholder participation using the cloud services e.g. formal channels for customers or agencies participating in any purchasing process. All participants in a cloud-service-based network need provide transparency and stick to a defendable process – no matter with which software this process is enabled.

## 8. APPLICATION OF THE APPROACH

The main master data are the tax payer data, tax, additional tax information and current tax situation. For processing the current challenge is that a lot of system interfaces / system breaks up to manual processing are in place. For the future the flexibility of various information channels are a must as well as one view towards the tax payer.

Tax and revenue departments must also address ever changing regulatory and tax law changes, and educate taxpayers about the impact of these changes. By creating automated workflows, any new system will enable the users to respond to taxpayer inquiries quickly and knowledgeably based on current information. Tax authorities gain greater visibility across revenue collecting activities and are better able to combat fraud and meet compliance mandates. Armed with complete information, civic leaders can estimate the impact of new taxes and fees and analyse the efficacy of new legislation as well as simplify debt disbursement management. By connecting fundamental business processes, tax authorities are able to provide accurate and clear representations of audit selection data and instigate predictive analytics to automatically identify taxpayer payment and filing irregularities. Integrating advanced data analytics can improve audit selection results and maximize revenue recovery efforts as well.

Here is an example how the security/compliance relevance could be quantified. We have explicitly chosen an example with a normalized value for the potential risk indicators. Whereas 1 indicates a very low level of damage/probability, 5 represents the maximal value of a very high level of damage/probability. The indicators we have chosen are very simple ones, yet they can easily be understood. E.g. „Interfaces" is an indicator of how probable a hacker attack might be, whereas „roles" indicates low likely internal fraud through access misconfiguration may be.

Table 1. Simplified risk indicators for the Order-to-Cash End-to-End process.

| Process Step Indicator | Speci-fication | Selection | Negotiation | Order | Fulfillment | Payment |
|---|---|---|---|---|---|---|
| Interfaces | 1 | 5 | 5 | 3 | 2 | 4 |
| Business relevance | 1 | 1 | 1 | 4 | 5 | 3 |
| Compliance requirements | 2 | 4 | 5 | 1 | 3 | 2 |
| Roles | 1 | 2 | 2 | 3 | 2 | 2 |
| Asset valuation | 1 | 2 | 2 | 2 | 4 | 5 |

Depending on the business model, the process, although the process steps are identical, might need us to adjust the indicators (or the weighting of their use, an option we will not discuss further in this paper) to reflect the impact of the business model choice (e.g. Push or Pull in Supply).

## 8.1. Security indicators and threats

The next step is to identify those processes and process steps that are subject to impact by the use of cloud services. This will address the security and compliance related indicators adequately. We propose to introduce a decision tree as a process reference. Such a decision tree allows us to identify those pieces of information that of major relevance in view of security and compliance requirements by breaking down the business process into sub-aspects that are specific for security and compliance, in the example in figure 3 given below called „GRC" (that stands for Governance, Risk and Compliance).

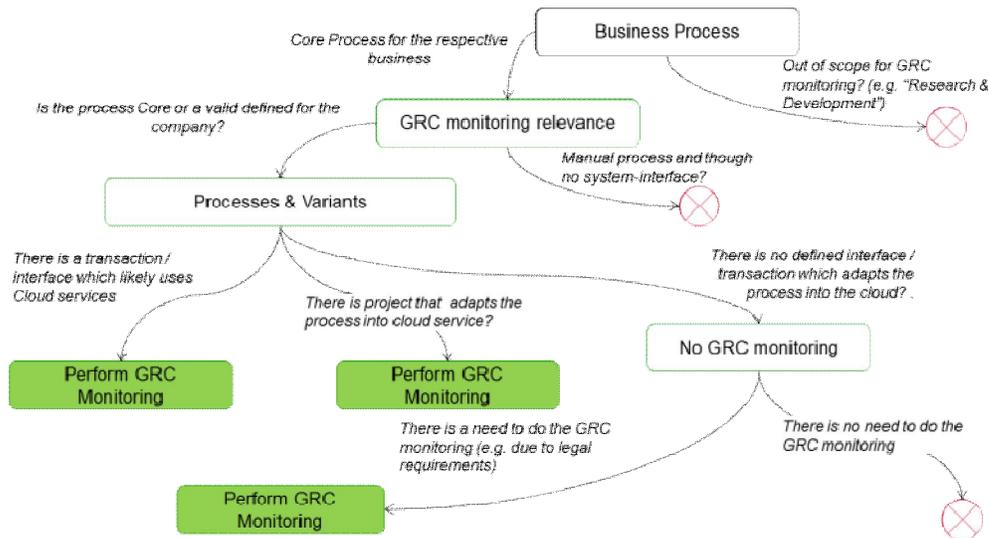

Figure 3: Decision tree for GRC activities

If we now apply this tool to the Order-to-Cash process, the security indicators are identified by comparing the conditions for information processing of the „classical" in-house service delivery and a cloud-based service delivery. To be able to do this, e.g. SAP-transactions used in-house must be mapped against corresponding cloud-based services that promise to fulfil the same process step. In table 2, we give an example for such a mapping, with corresponding attributes and resulting risks, for one transaction / cloud service.

Table 2. Risk indicators in comparison for in-house and cloud-based delivery - an example.

|  | SAP transaction ME21N (create order) | Corresponding cloud service | Resulting risk of moving to the cloud |
|---|---|---|---|
| Interfaces | 2 | 5 | SIGNIFICANTLY HIGHER |
| Business relevance | 3 | 3 | NO ADDITIONAL RISK |
| Compliance requirements | 3 | 3 | NO ADDITIONAL RISK |
| Roles | 4 | 3 | LOWER |
| Asset valuation | 2 | 2 | NO ADDITIONAL RISK |

## 9. CONCLUSIONS

We have shown in this paper how to choose Public Sector business process elements that are suitable to be transferred to the cloud. To achieve this, we have elaborated a value-chain driven approach to attribute specific process elements with benefits and risks. It is especially important to look at the end-to-end process chains to capture the full value proposition, even if they cross organizational borders - something that is typical for Public Sector processes. For the individual valuation of the process steps we have given examples, further work is needed to fully understand which aspects are invariant and which aspects need to be investigated in a real organizational context.

**Authors**

Sachar Paulus is Professor for Information Systems and Security Management at Brandenburg University of Applied Sciences since 2009. Prior to this, he was active in the IT and Security industry, and among others, Chief Security Officer of SAP. He has a Ph.D. in mathematics.

Ute Riemann is Principal Business Consultant with a focus on value-chain orientation at SAP since 2012. Prior to this, she was active in numerous positions related to process architecture. She has an MBA and diplomas in Computer Science and Business Administration.